\pdfoutput=1
\documentclass[preprint,twocolumn,tightenlines,10pt,prl]{revtex4}%
\usepackage{amsfonts}
\usepackage{amsmath}
\usepackage{amssymb}
\usepackage{graphicx}%
\providecommand{\U}[1]{\protect\rule{.1in}{.1in}}

\begin{document}
\title{A Dynamical Crossover Regime in the Transmission and Reflection Spectra of
Evanescent Waves with 2D Arrays of Josephson Junctions}
\author{L. L. A. Adams$^{\ast}$}
\affiliation{The Department of Physics and School of Engineering and Applied Sciences,
Harvard University, Cambridge, MA 02138}


\begin{abstract}
A dynamical cross-over regime is revealed when exposing a classical
two-dimensional ordered Josephson junction (JJ) array to evanescent waves and
tuning the incident microwave power. At the lowest possible temperature for
these experiments, 1.1 K, and at the lowest power setting, -55 dBm, evanescent
waves are transmitted without loss and the resonance exhibits a quality factor
of \symbol{126}4200. A second, smaller resonance, which evolves with
increasing power from the main resonance, is also investigated. In contrast to
the behavior of the main resonance, this second peak grows as the incident
power is increased and does not maintain a fixed resonant frequency for
temperatures less than the superconducting critical temperature of niobium.
The tunability of both resonances is studied as a function of temperature and
microwave power. Finally we speculate that this dynamical crossover regime is
evidence of a transition between two states of phase coherence where at low
microwave power the JJ arrays are phase locked and at high microwave power the
JJ arrays are unlocked.\TeX{} .

\end{abstract}
\volumeyear{year}
\volumenumber{number}
\issuenumber{number}
\received[Received text]{date}

\revised[Revised text]{}

\startpage{1}
\endpage{102}
\maketitle

Negative index of refraction metamaterials, materials that simultaneously have
a negative permittivity and negative permability, bend electromagnetic waves
in counterintuitive directions.\cite{Veselago} One promise of these materials
is their ability to transform decaying evanescent waves into amplifying waves
and, by doing so, extend the resolution of optical microscopes to
sub-wavelength range. \cite{pendry1} However, this opportunity \cite{Zhang} is
jeopardized by dissipative losses which grow as the size of the constituents,
the meta-atoms, are made smaller.\cite{Smith} An encouraging remedy to
overcome losses is to directly embed gain media into metamaterials having
fishnet geometries. Theoretical models suggest that as a result of loss
compensation, electromagnetic signals can be amplified with 'high' quality
factors, as 'high' as 800, for realistic values of gain.\cite{Yang} While
amplifying\textit{ propagating} electromagnetic waves has successfully been
demonstrated, albeit with signals having a low quality factor, Q
$<$
50, \cite{Wuestner} a similar, tunable amplification of \textit{non}%
-propagating evanescent waves with high quality factors remains to be seen.
And since losses and lack of tunability are the real show stoppers for future
applications, achieving the remarkable properties known to metamaterials
demands removing nearly all losses \cite{Ricci} while embedding nonlinear
elements for tunability.

Lossless materials which combine fishnet, or square loop, geometries with
superconducting nonlinear gain media are known as two dimensional ordered
Josephson junction\ arrays. Josephson junction (JJ) arrays have intrinsic
capacitances and inductances built into them - a key component in the design
of negative index of refraction metamaterials. As attractive tunable
metamaterials their salient features are their low loss nature, nonlinear
inductance, and miniaturized device elements.\cite{Zheludev} All of this leads
to materials that are well suited as metamaterials but in practicality, JJ
arrays, at least conventional JJ arrays, would not be of convenience since
they need to be kept cold and radiated with microwaves. Nevertheless, if this
is done, then these are ideal materials for restoring decaying evanescent waves.

In this Letter, we report low temperature measurements of evanescent wave
amplification with quality factors as high as 4200 when using unbiased
superconducting JJ arrays. Evanescent wave amplification, as described \ in
the transmission experiments of Ref \cite{Baena}, is the sharp increase in
transmission through a waveguide with a metamaterial in the same frequency
range as the cutoff regime of an empty waveguide. We also observe tunability
and a surprising transition between two different sets of resonances with
drastically different responses to temperature and incident power. Once the
temperature exceeds the superconducting transition temperature, the two
tunable resonances no longer prevail suggesting their relevance to the
sample's superconducting properties.

The samples are two dimensional JJ arrays that were commercially fabricated by
the Sperry-Univac Corporation, now Unisys, using the selective niobium
annodization process \cite{Kroger} and constructed on a silicon substrate
covered with a 1 $\mu m$ thick oxide layer. \cite{Goldman} These arrays,
arranged in a square ring geometry are unique in that there is no ground
plane. Each ring has four equally spaced Josephson junctions and is defined
here as a "meta-atom" whose size is much smaller than the applied microwave
wavelength, thus creating an effective medium. The length of one side is 29
microns, with a linewidth of 5 microns and a niobium thickness of 500 nm. The
junctions have an area of (2.5 microns)$^{2}$ and are made of Nb-amphorous
Si-Nb, with a-Si serving as a 10 nm thick insulating barrier.\ A\ portion of
the JJ arrays is shown in Fig.1\thinspace a. The critical current of an
individual junction is estimated to be 1.65 x 10$^{-7}$ A. Transport
measurements in a magnetic field on these arrays are reported elsewhere.
\cite{Elsayed} The superconducting transition temperature is T$_{c}=\,9.2\,K$
as measured by SQUID$\ $magnetometry.

The original sample with one million (1000 x 1000) junctions is divided into
several pieces with each piece containing only the arrays and\textit{ not}
electrodes or pads for electrical contacts. The samples are placed in a
Ag-plated Cu X-band rectangular waveguide and held in place with a piece of
Rohacell, a styrofoam-like material with a dielectric constant between 1.06 -
1.11 and oriented such that the rf magnetic field is perpendicular to the
surface of the sample.

Attached to the rectangular waveguide are heaters, a thermometer and two
antennas which are housed in a vacuum can inside a magnetically shielded
cryostat. Semi-rigid coaxial cables \cite{RFDepot} connect to the antennas and
extend to a HP 8722 D vector network analyzer via phase-maintaining coaxial
cables. \cite{Megaphase} Standard transmission and reflection measurements are
made by acquring scattering parameters; the data is averaged over 16
subsequent runs with 1601 data points taken for each run. No additional
filters or amplifiers are added to the system. The cutoff frequency for an
empty waveguide is 6.56 GHz as shown in Fig. 1b.%
\begin{figure}
[ptb]
\begin{center}
\includegraphics[
natheight=7.027500in,
natwidth=3.499900in,
height=4.28in,
width=2.1508in
]%
{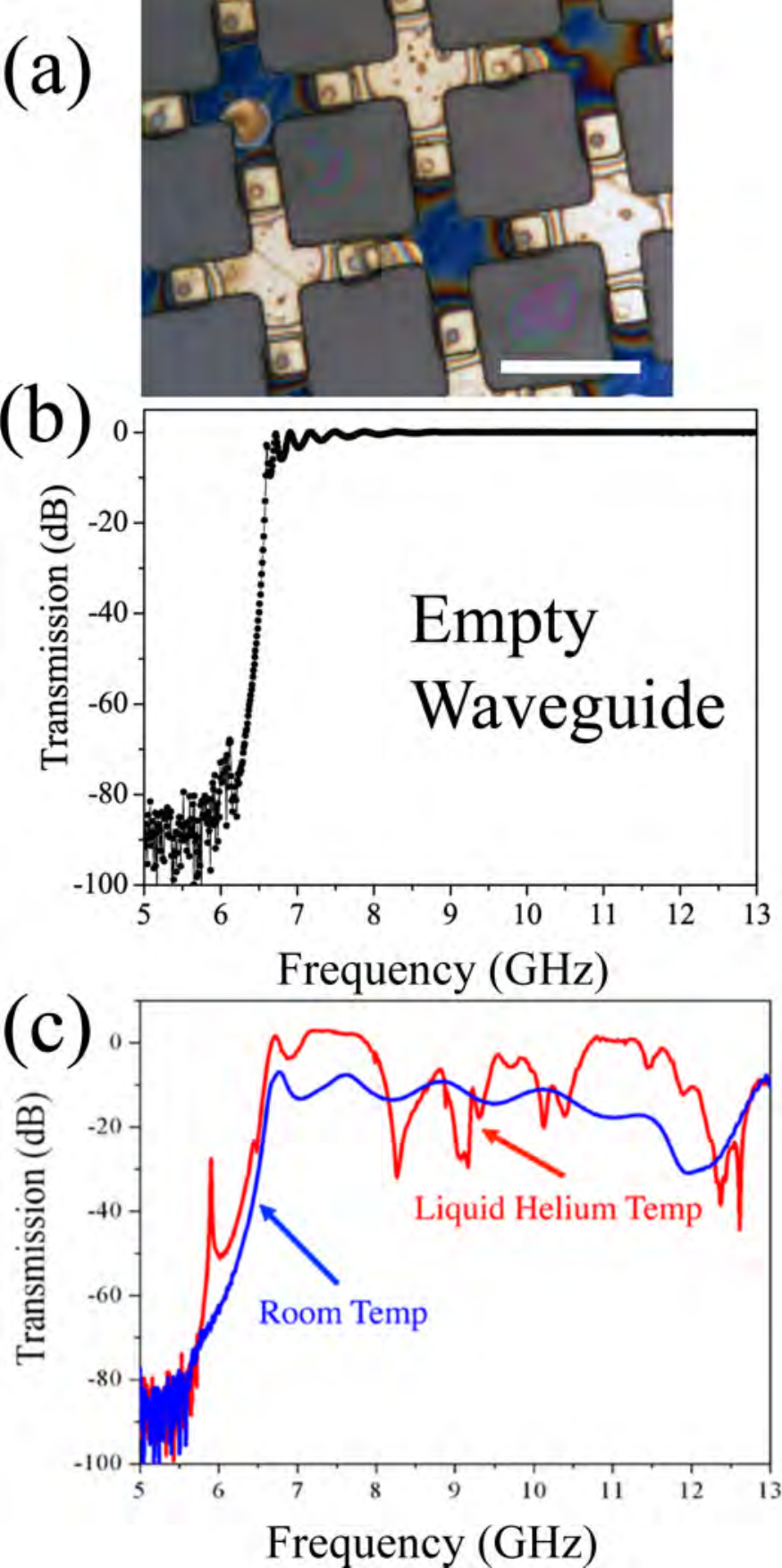}%
\caption{(color online) (a) Optical microscope image of \ a portion of the
sample. Each square loop contains 4 Josephson junctions; the scale bar denotes
30 $\mu m.$(b) Plot of transmission vs frequency for an empty waveguide. (c)
Plot of transmission vs frequency of the same waveguide as above but
containing several unbiased samples at room (blue curve) and liquid helium
(red curve) temperatures.Note: The value of +3 dB between 7-8 GHz for the
liquid helium curve corresponds to 0 dB; this offset is due to the calibration
being done at room temperature while the measurements are made at liquid
helium temperatures.}%
\end{center}
\end{figure}

We place several small, unbiased samples of different sizes in the waveguide.
At room temperature, well above T$_{c}$ of the JJ arrays, the electromagnetic
transmission has broad ripples and a broad resonance at 12 GHz as shown in the
blue curve of Fig. 1c. Additionally, there is noticeable insertion loss of -7
dB and the signal from 5 GHz to the cutoff frequency is much broader than for
the empty waveguide. By cooling to liquid helium temperatures, where the
arrays are superconducting, several sharp resonances appear at discrete
frequencies both above and below cutoff as shown in the red curve of Fig. 1c.
Also, the insertion loss goes to zero indicating that the lossy spectrum at
room temperature is a consequence of the normal state JJ arrays and not the
substrate. \ 

To explore these resonant features in greater detail and without the
possibility of crosstalk, we place a single sample of 0.5 cm x 0.5 cm
dimensions with roughly 30,000 meta-atoms at the center of the waveguide. At
fixed temperature, T = 1.1 K, two distinctive transmission features appear
below cutoff with increasing incident power from -55 dBm to - 5 dBm having
drastically different behaviors: the dominant resonant peak \textit{decreases}
in amplitude at fixed frequency then slightly shifts to lower frequencies over
a range of $\thicksim$ 0.4 MHz, until a second, smaller resonance develops and
\textit{increases} in amplitude with increasing frequency over a range of
$\thicksim$13 MHz as shown in Fig. 2a. Increasing the power also has the
effect of broadening the dominant resonance, Q decreases from 4200 to 340,
while sharpening the smaller resonance, Q increases from 340 to 1240. \
\begin{figure}
[ptb]
\begin{center}
\includegraphics[
natheight=5.430200in,
natwidth=3.499900in,
height=3.845in,
width=2.4898in
]%
{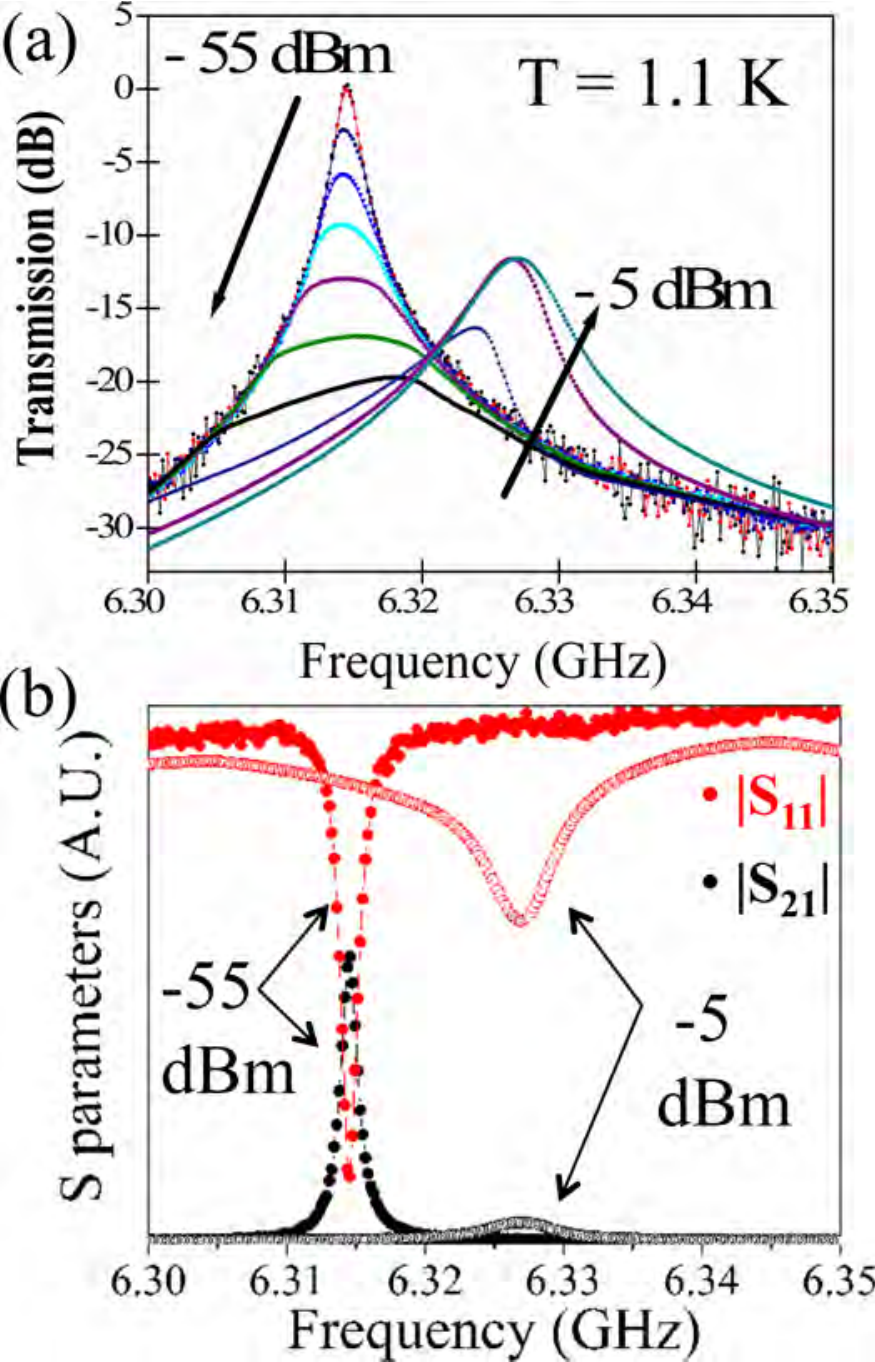}%
\caption{(color online) Measurements at T = 1.1 K. (a) Frequency dependence of
array transmission with input power ranging from -55 dBm to -5 dBm, every -5
dBm. The arrows indicate increasing input power. (b) \ Magnitude of the
transmission $\shortmid$ S$_{21}\shortmid$ (black circles) and reflection
$\shortmid$ S$_{11}\shortmid$ \ (red circles) at -55 dBm and - 5 dBm.
\cite{adams}}%
\end{center}
\end{figure}
The amplitude of the dominant resonance is tunable over 30 dBm with a "net
gain" of 20 dB, and the smaller resonance is tunable over 10 dBm with a "net
gain" of 8 dB. Here "net gain" refers to the increase in the signal's
amplitude from negative dB to 0 dB; this is \textit{not a real gain} since the
signal never exceeds 0 dB.

The observed transmission of evanescent waves from one end of the waveguide to
the other occurs by tunneling since evanescent waves cannot carry energy.\ The
signature of `transmission by tunneling' is equal, or near equal, amplitudes
of forward and backward evanescent modes \cite{Baena},\cite{pendry2}; this is
shown in the reflection and transmission coefficients and plotted as a
function of frequency for -55 dBm and -5 dBm in Fig. 2b.\cite{adams}

The resonant frequency, \textit{f}, of a single JJ meta-atom is estimated from
the geometric inductance, \textit{L}$_{g},$ the Josephson inductance,
\textit{L}$_{J},$ and the capacitance, \textit{C,} of the tunnel junctions.
This is expressed as $f\!=\frac{1}{2\pi}\!\frac{1}{\sqrt{\left(  \frac
{4L_{J}L_{g}}{4L_{J}+L_{g}}\right)  \times4C}}$ \ with \textit{L}$_{g}=\mu
_{o}R\left[  \ln\left(  \frac{8R}{r}\right)  -\frac{7}{4}\right]  $ where R is
the radius of the square loop and r is the radius of the wire. \cite{Vier}
\ From the dimensions of an JJ meta-atom\ and the Josephson inductance \ which
is defined as L$_{J}=\frac{\Phi_{o}}{2\pi I_{c}}$with $\Phi_{o}=\frac{h}{2e}$
where \textit{I}$_{c}$ is the critical current, \textit{h }is Planck's
constant, and \textit{e} is the charge of the electron, the estimated resonant
frequency is $\thicksim$ 50 GHz. This crude estimate is an order of magnitude
higher than the observed resonant frequency suggesting that additional terms
have been neglected, such as the kinetic inductance or possibly the estimated
critical current is incorrect. In addition, this calculation is not accurate
in describing the experimental situation since the resonant frequency depends
on the location and orientation of the sample inside the waveguide. We compare
this calculation to the Josephson plasma frequency\bigskip, $f\!=\frac{1}%
{2\pi}\!\sqrt{\frac{2eI_{c}}{\hbar C}}\thicksim$ 15 GHz, which is closer to
the actual value.

Given the discrepancies between predicted values and experimental
observations, it is necessary to relate the observations to superconductivity
and the Josephson effect. For superconductivity, microwave measurements at
different temperatures are made, with four different temperatures shown in
color plots of Fig. 3. Perfect transmission, 0 dB, at T =1.1 K and 6.0 K for
-55 dBm is shown in green in Figure 3 a and b.

Interestingly, as the temperature approaches T= 6.0 K, higher and higher input
powers are required for the emergence of the second peak; this behavior is
clearly seen when comparing the transmission spectrum in Fig. 3b to that in
Fig. 3a. One possible explanation for this is that with increasing
temperature, until T=6.0 K, more junctions become coherent with each other as
larger numbers of their individual linewidths overlap. Thermally induced
coherence can be a result of the sample not being precisely centered in the
waveguide or perfectly oriented with the rf magnetic field. The onset of
coherence with temperature was\ previously mentioned in Ref. \cite{Wahlsten}.

This situation drastically changes for T
$>$
6.0 K: the tunability of each resonance substantially decreases with
increasing power. Indeed, since these are two dimensional JJ arrays, they are
model systems for the Berezinskii-Kosterlitz-Thouless (BKT) phase transition.
As previously reported for these samples, the BKT transition temperature
between the binding of vortex and anti-vortex pairs (T%
$<$%
T$_{BKT}$), and the unbinding of these pairs (T%
$>$%
T$_{BKT}$) occurs at T$_{BKT}$ = 5.85 K. \cite{Elsayed} Our data of a turning
point at T= 6.0 K is consistent with this picture. At T = 8.0 K, the
transmission of the smaller, higher frequency resonance saturates and is no
longer tunable with increasing power as shown in Fig. 3c. However, the main
resonance is still somewhat tunable at T=8.0 K, although transmission occurs
at lower frequencies than the resonances at T $\preceq$ 6.0 K. \ At T= 10 K
and beyond, there is only one broad, low amplitude resonance which is
constant, albeit noisy, with incident power as shown in Fig. 3d; in aggregate,
these temperature and incident power results are indicative of
superconductivity.
\begin{figure}
[ptb]
\begin{center}
\includegraphics[
natheight=4.330100in,
natwidth=3.499900in,
height=3.7766in,
width=3.0588in
]%
{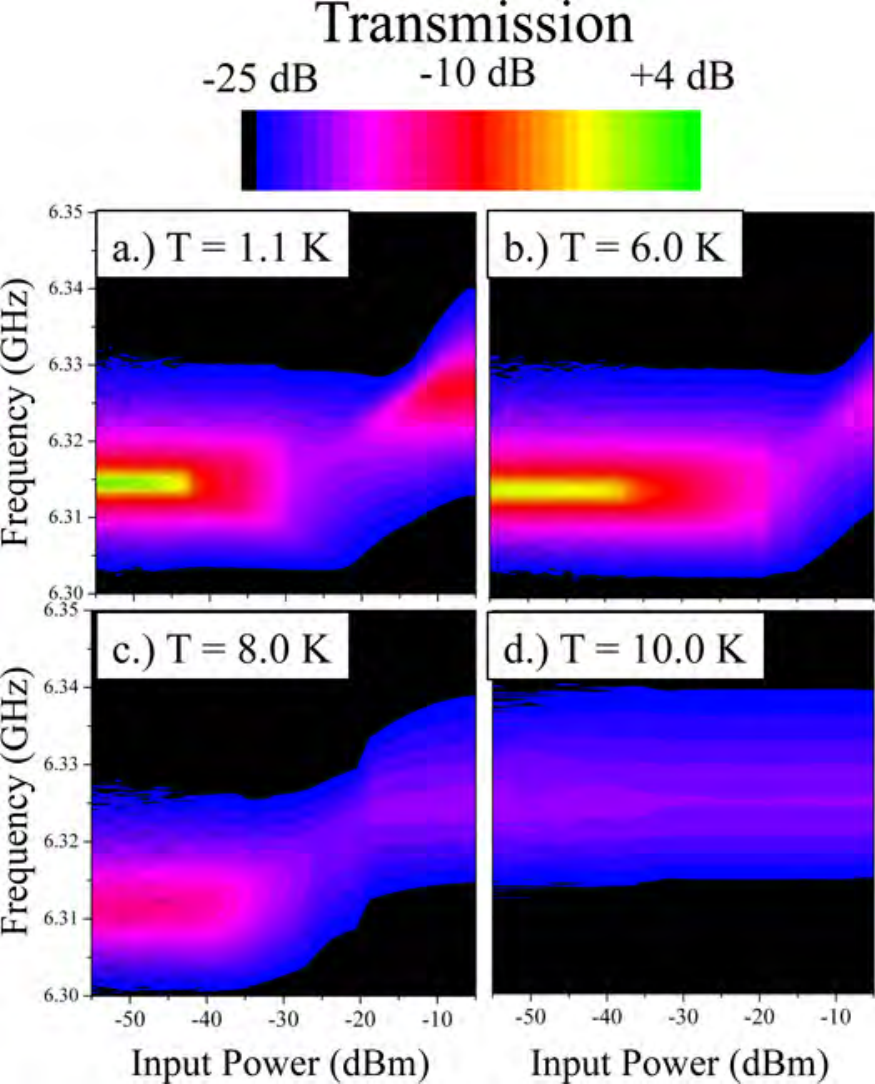}%
\caption{(color online). Microwave transmission (dB) as a function of
frequency and input power for a.) T=1.1 K, b.) T = 6.0 K, c.) T = 8.0 K, and
\ d.) T =10.0 K. The color bar scale represents the range of transmission from
zero transmission (black) to perfect transmission (green). Note: The value of
+4 dB corresponds to 0 dB due to the calibration being done at room
temperature while the measurements are made at liquid helium temperatures.}%
\end{center}
\end{figure}

Central to understanding the nonlinearity and Josephson effect of these
spectra requires explaining the tunability of the 6.314 GHz resonance for T
$\preceq$ 6.0K. If the reduction in amplitude with increasing power is a
result of dissipation, then as T increases, the amplitude should decrease.
Keeping power fixed at -55 dBm and increasing T, the amplitude decreases.
However, the decrease in amplitude at fixed power with temperature is not of
the same ilk as the decrease in amplitude at fixed T with increasing power as
seen in Fig. 2a and Fig. 3 . First, as T increases at fixed power, the
resonant frequency shifts to lower values than for the case of increasing
power at fixed T as shown in Fig. 4a. Secondly, and rather remarkably,
tunability of the resonance occurs with very low incident power even though
the niobium thickness is 500 nm. For even thinner films of niobium,
\symbol{126}200 nm, without JJs, significantly higher values of incident power
are necessary, as high \symbol{126}+20 dBm, to lower the amplitude by
\symbol{126}20 dB; for thin Nb films, this amplitude reduction, accompanied
\ with a downward frequency shift, is due to dissipative losses. \cite{Kurter}
Lastly, and most surprisingly, is the decrease in loss, 1/Q, \ with increasing
T for input powers between -40 dBm and \symbol{126}\thinspace\ -18 dBm for T
$<$
6 K. This is shown in the shaded region of Fig. 4b and confirms that
dissipation is not driving the decrease in amplitude with increasing input
power of the main resonance.
\begin{figure}
[ptb]
\begin{center}
\includegraphics[
natheight=6.582900in,
natwidth=3.499900in,
height=4.5602in,
width=2.4431in
]%
{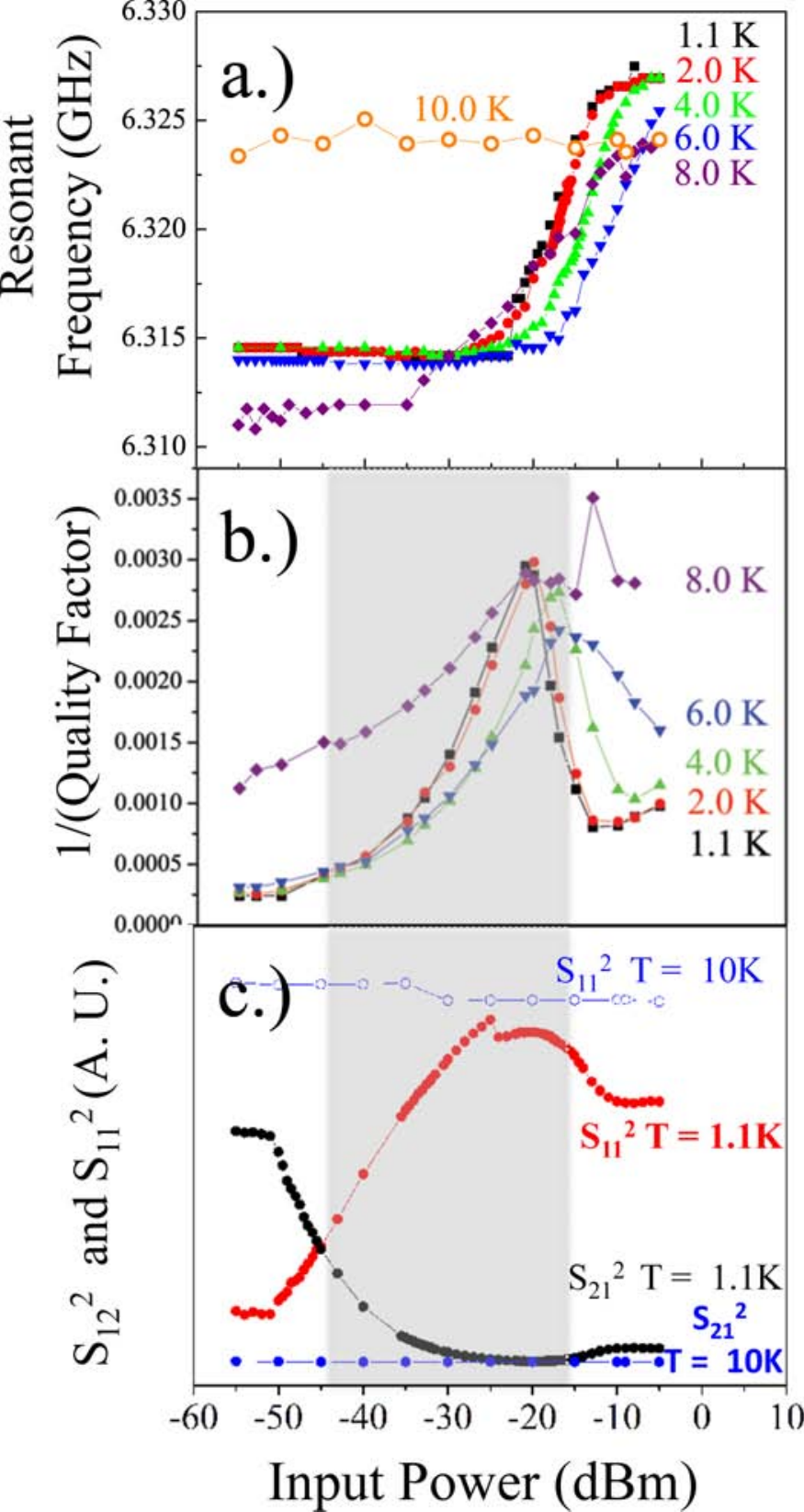}%
\caption{(color online). (a) Resonant frequency \ vs input power for different
temperatures. (b) (1/Quality Factor) vs input power for different
temperatures. The shaded gray region, -40 dBm to -18 dBm, indicates a regime
where 1/Q decreases with increasing temperature from 1.1 K to 6.0 K. (c)
S$_{12}^{2}$ and S$_{11}^{2}$ vs input power for T=1.1 K and T = 10 K.}%
\end{center}
\end{figure}

If the tunability of the main resonance is not due to dissipation, then
another possibility is that the JJs are operating as parametric amplifiers
owing to their nonlinear inductance. The similarity between these results and
those due to parametric amplification might not be coincidental since the
signal from the network analyser is not a pure tone, but rather contains
sidebands. It is by these sidebands that we are likely seeing "gain". We
measured the output signal at 6.4 GHz from the network analyzer; there are
sidebands spaced at +/- 20 kHz and +/- 50kHz as well as higher order
harmonics. In addition to a central frequency, sidebands from separate sources
are standard for amplifying signals with JJ arrays; however, in these
experiments, the amplitude of the signal is always greater than 0
dB.\cite{Zimmer} - \cite{Haviland}

The losses, 1/Q, increase with increasing T for high power resonances, greater
than -18 dBm, indicating increasing dissipation with temperature as shown in
Figure 4 b. Additionally, since these resonances shift to higher frequencies
with incident power as shown in Figs. 2 a \& 4 a this suggests that these
resonances are not due to quasiparticles, but likely free vortices. When
comparing 1/Q to the transmission and reflection coefficients we observe that
as 1/Q increases or decreases, the reflection coefficients follow suit as
shown for T = 1.1 K in the red curve in Fig. 4c. Above the superconducting
transition temperature, T = 10 K, the signal is all reflection as shown in the
open circles of the blue curve of Fig. 4c. So,why then, does one peak evolve
into a second peak? Why are there two peaks? What is the nature of the second
peak? These are left as open questions and require further investigation.

However, and without a corroborative theory to back up our findings,
we\ naively speculate that the difference between these two sets of resonances
is a measure of how phase sensitive the JJ arrays are to incident power: we
speculate that the arrays are phase insensitive at low power and thus are
phase locked for the dominant resonances, while for the second set of
resonances, at higher incident powers, the arrays are phase sensitive, and
thus unlocked. This results in a constant resonant frequency for the phase
locked arrays, but a shifting resonant frequency for the unlocked arrays with
incident power.

We also base our thinking on eliminating what these resonances are not: first,
we rule out generation of quasiparticles with increasing incident power for T
$\preceq$ 6.0 K \ because in our work increasing power shifts the resonances
to higher frequencies not lower frequencies as would otherwise be the case.
Second, we rule out extraordinary transmission \cite{Ebbesen} since the
resonant frequency and sample location in the waveguide are coupled to each other.

In conclusion, tunable amplification of evanescent waves has been demonstrated
using unbiased, fishnet Josephson junction metamaterials with a simple
experimental setup that does not require additional filtering, amplifiers or
biasing of the sample. This experimental simplicity together with the
robustness and tunability of the evanescent signal over a wide range of
incident powers and temperatures make these experimental results all that more
compelling. Most noteworthy, however, is the unanticipated transition from one
set of resonances to a second set of resonances. This transition could be a
manifestation of the arrays transitioning from being locked in phase to being
unlocked; however, further investigations are needed to confirm this
speculation or provide an alternative explanation.

Acknowledgements. The author gratefully acknowledges the assistance of her
advisor, S. Anlage, and R. Anderson, K. Lehnert, B. Yurke, C. Lobb, J. Koch,
J. Hamilton, M. Devoret, S. Girvin, R. Jaramillo, D. Stroud, C. Kurter and
T.Kodger. The author also thanks A. Goldman for providing samples. This work
was supported by the Intelligence Community Postdoctoral Research program and
carried out at the University of Maryland, Center for Nanophysics and Advanced
Materials, College Park, MD.

*Email address: lladams@seas.harvard.edu

\end{document}